\documentstyle[preprint,aps]{revtex}
\begin{document}
%
%
\preprint{$
\begin{array}{l}
\mbox{ILL-(TH)-99-4}\\[-3mm]
\mbox{UB-HET-99-01}\\[-3mm]
\mbox{September~1999} \\ [1cm]
\end{array}
$}
\title{Two Photon Radiation in $W$ and $Z$ Boson Production at the
Tevatron Collider\\[1cm]}
\author{U.~Baur\footnote{e-mail: baur@ubhex.physics.buffalo.edu}}
\address{Department of Physics,
State University of New York, Buffalo, NY 14260, USA\\[3.mm]}
\author{T.~Stelzer\footnote{e-mail: tstelzer@uiuc.edu}}
\address{Department of Physics,
University of Illinois, 1110 West Green Street, Urbana, IL 61801,
USA\\[10.mm]}
\maketitle
%
%
%
\begin{abstract}
\baselineskip15.pt  
We present a calculation of two photon radiation in $W$ and $Z$ boson
production in hadronic collisions, based on the complete matrix
elements for the processes $q\bar q'\to\ell^\pm\nu\gamma\gamma$ and
$q\bar q\to\ell^+\ell^-\gamma\gamma$, including finite charged lepton
masses. In order to achieve stable numerical results over the full phase 
space, multiconfiguration Monte Carlo techniques are used to map the
peaks in the differential cross section. Numerical results are presented 
for the Fermilab Tevatron. 
\end{abstract}
\newpage
%
%
\section{Introduction}
The Standard Model of electroweak interactions (SM)
so far has met all experimental challenges and is now tested at the 
$0.1\%$ level~\cite{holliksm}. However, there is little direct
experimental information on the mechanism  which generates the masses of
the weak gauge bosons. In the SM, spontaneous symmetry breaking is
responsible for mass generation. The existence of a Higgs boson
is a direct consequence of this mechanism. At present the negative
result of direct searches performed at LEP2 imposes a lower bound
of $M_H>98.8$~GeV~\cite{ewwg} on the Higgs boson mass. Indirect
information on the mass of the Higgs boson can be extracted from the 
$M_H$ dependence of radiative corrections to the $W$ boson mass, $M_W$,
and the effective weak mixing angle, $\sin^2\theta^{lept}_{eff}$.
Assuming the SM to be valid, a global fit to
all available electroweak precision data yields a (one-sided) 95\%
confidence level (CL) upper limit on $M_H$ of about
260~GeV~\cite{holliksm,ewwg,caso}. 

Future more precise measurements of $M_W$ and the top quark
mass, $m_{top}$, will lead to more accurate information on the Higgs
boson mass~\cite{degrassi,Tev2000,BD}. Currently, the $W$ boson mass is 
known to $\pm 42$~MeV~\cite{lanc} from direct measurements. The
uncertainties of the individual experiments contributing to this value
are between about 80~MeV and 110~MeV~\cite{lanc,wnote}. The present 
uncertainty of the top quark mass from direct measurements 
is $\pm 5.1$~GeV~\cite{ward}. With a precision
of 30~MeV (10~MeV) for the $W$ mass, and 2~GeV for the top quark mass, 
$M_H$ can be predicted from a global analysis with an uncertainty of
about $30\%$ ($15\%$)~\cite{Tev2000,BD}. Comparison of these indirect 
constraints
on $M_H$ with the results from direct Higgs boson searches at LEP2, the 
Tevatron collider, and the Large Hadron Collider (LHC) will be an
important test of the SM. They will also provide restrictions on the
parameters of the Minimal Supersymmetric extension
of the Standard Model (MSSM)~\cite{hollikmssm}.

A significant improvement in the $W$ mass uncertainty is expected in the
near future from measurements at LEP2~\cite{LEPWmass} and the Fermilab
Tevatron $p\bar p$ collider~\cite{Tev2000}. The ultimate precision
expected for $M_W$ from the combined LEP2 experiments is  
30~-- 40~MeV~\cite{LEPWmass}. At the Tevatron, integrated luminosities of
order 2~fb$^{-1}$ are envisioned in the Main Injector Era (Run~II), and one
expects to measure the $W$ mass with a precision of approximately
40~MeV~\cite{Tev2000} per experiment. The prospects for a precise 
measurement of $M_W$ would further improve if a significant upgrade in 
luminosity beyond the goal of the Main Injector could be realized. 
With recent advances in
accelerator technology~\cite{GPJ}, Tevatron collider luminosities of
order $10^{33}\,{\rm cm}^{-2}\,{\rm s}^{-1}$ may become a reality,
resulting in integrated luminosities of up to
10~fb$^{-1}$ per year. With a total integrated luminosity of
30~fb$^{-1}$, one can target a precision of the $W$ mass of 15~--
20~MeV~\cite{Tev2000}. A similar or better accuracy may also be reached 
at the LHC~\cite{KW}.

In order to measure the $W$ boson mass with high
precision in a hadron collider environment, it is necessary to fully 
understand and control higher order QCD and electroweak (EW) corrections 
to $W$ production. The determination of the $W$ mass in a hadron 
collider environment
requires a simultaneous precision measurement of the $Z$ boson mass,
$M_Z$, and width, $\Gamma_Z$. These quantities serve as reference
points. When compared to the value measured at LEP, 
they help to accurately determine the energy scale and
resolution of the electromagnetic calorimeter, and to constrain the
muon momentum resolution~\cite{Tev2000}. In order to extract $M_W$ from
hadron collider data, it is therefore also necessary to
understand the higher order QCD and EW corrections to $Z$ boson 
production in hadronic collisions. 

Electroweak radiative corrections have a significant impact on the $W$
and $Z$ boson masses and widths extracted from experiment. Recent 
improved calculations of the
${\cal O}(\alpha)$ EW corrections to $W$ production~\cite{BKW}, 
and of the ${\cal O}(\alpha)$ QED corrections to $Z$ production in
hadronic collisions~\cite{BKS}, have shown that the main effect is caused 
by final state photon radiation. When detector effects are included, 
${\cal O}(\alpha)$ radiative corrections shift the $W$ mass by about 
$-50$~MeV in the electron case, and approximately $-160$~MeV in the muon 
case~\cite{cdfwmass,D0Wmass}. The effect on the $Z$ mass is about a
factor two larger than that on $M_W$ for both electron and muon
final states. ${\cal O}(\alpha)$ photon emission also shifts the width
of the $W$ boson extracted from the tail of the transverse
mass distribution by approximately $-70$~MeV~\cite{wwidth}. The size of 
the shift in $M_W$, $M_Z$ and the $W$ width introduced by the
${\cal O}(\alpha)$ corrections raises the question of how strongly
${\cal O}(\alpha^2)$ corrections affect these quantities.

In order to reliably calculate the impact of the 
${\cal O}(\alpha^2)$ corrections to 
$p\,p\hskip-7pt\hbox{$^{^{(\!-\!)}}$} \to W^\pm\to\ell^\pm\nu$ and
$p\,p\hskip-7pt\hbox{$^{^{(\!-\!)}}$} \to \gamma^*, Z\to\ell^+\ell^-$ 
($\ell=e,\, \mu$) on the $W$ and $Z$ masses extracted from experiment,
a full calculation including real and virtual
corrections, which is valid over the entire allowed phase space, is
needed. So far, only partial calculations for the ${\cal 
O}(\alpha^2)$ real photon corrections, $p\,p\hskip-7pt\hbox{$^{^{(\!-\!)
}}$} \to W^\pm\to\ell^\pm\nu\gamma\gamma$ and $p\,
p\hskip-7pt\hbox{$^{^{(\!-\!)}}$} \to \gamma^*,
Z\to\ell^+\ell^-\gamma\gamma$ exist~\cite{photos,BHKSZ}. In
Ref.~\cite{photos} the structure function approach for photon radiation
is used to perform the calculation and, therefore, only final state 
photon radiation in the leading log approximation 
is included. The results obtained using
this approach are reliable only for small opening angles between
the photons and the charged leptons. The calculation of 
Ref.~\cite{BHKSZ} is based on the full set of tree level Feynman diagrams
contributing to $\ell\nu\gamma\gamma$ and $\ell^+\ell^-\gamma\gamma$
production. In addition, to preserve gauge invariance when finite $W$ width
effects are included, the imaginary part of the $WW\gamma$ and
$WW\gamma\gamma$ one-loop vertex corrections is taken into account. 
However, charged leptons are assumed to be massless, and thus
a finite lepton -- photon separation cut has to be imposed in order to
avoid the collinear singularities associated with final state radiation.

The first step towards a calculation of the ${\cal O}(\alpha^2)$
corrections to $W$ and $Z$ boson production in hadronic collisions thus 
is to perform a calculation of $\ell\nu\gamma\gamma$ and
$\ell^+\ell^-\gamma\gamma$ production which 
\begin{itemize}
\item is based on the full set of
Feynman diagrams contributing at tree level, 
\item includes finite lepton
mass effects, 
\item is gauge invariant when finite $W$ width effects are taken
into account, 
\item and is valid for arbitrary lepton -- photon opening 
angles. 
\end{itemize}
In addition, in order to obtain reliable information
on the shift in $M_W$ and $M_Z$ caused by two photon radiation, the
numerical calculation should be stable for photon energies as small as
the tower threshold of the electromagnetic calorimeter of the Tevatron
experiments, which is of ${\cal O}(100$~MeV). In this paper we present such a
calculation. 

While the calculation of the $q\bar q\to\ell^+\ell^-\gamma\gamma$ and $q\bar
q'\to\ell\nu\gamma\gamma$ matrix elements is straightforward, the phase 
space integration presents some challenges, due to the sharp peaks
in the matrix elements which arise from the soft and collinear
singularities. The collinear singularities associated with final state
radiation are regulated by the finite mass of the leptons whereas soft and
initial state collinear singularities are rendered finite by transverse 
momentum cuts imposed on the photons. Both soft and collinear singularities 
produce large
contributions to the cross section in small regions of phase space.
Standard adaptive Monte Carlo integration routines such as
VEGAS~\cite{vegas} do not yield a numerically stable result of the cross
section for processes which exhibit a complicated peaking structure in
the matrix elements. To obtain numerically stable and accurate results
for such processes, a multi-channel Monte Carlo approach~\cite{excal}, 
augmented by the adaptive weight optimization procedure described in 
Ref.~\cite{KP}, is frequently used. The disadvantage of the multi-channel
Monte Carlo approach is that the peaks in the matrix elements have to be
mapped by hand, thus requiring a substantial amount of analytic work 
which has to be repeated for each new process one wishes to analyze. 

Our calculation is based on a similar approach which adds the benefit of
largely automizing the mapping of the peaks in the matrix elements. The 
process independent features of our approach, and the resulting 
multiconfiguration Monte Carlo (MCMC) integration program, are briefly 
described in 
Sec.~II. Full details will be given elsewhere~\cite{tim}. In Sec.~III we
discuss technical details associated with the calculation of the 
$q\bar q\to\ell^+\ell^-\gamma\gamma$ and $q\bar
q'\to\ell\nu\gamma\gamma$ matrix elements and present 
numerical results for two photon radiation in $W$ and $Z$ events at the
Tevatron collider ($p\bar p$ collisions at 1.8~TeV). Finally, 
summary remarks are given in Sec.~IV.

\section{Phase Space Integration}

The matrix elements for $q\bar q\to\ell^+\ell^-\gamma\gamma$ and $q\bar
q'\to\ell\nu\gamma\gamma$ have many sharply peaked regions
throughout phase space. In addition to the Breit-Wigner resonances
around the $W$ or $Z$ pole and a pole at small $\ell^+\ell^-$ invariant
masses due to photon exchange, there are singularities when either 
photon becomes soft, or collinear with a charged particle. Although these 
soft and collinear singularities are regulated by energy or transverse 
momentum cuts and fermion masses, they result in large
contributions to the cross section over relatively small regions of
phase space and cause difficulties for standard integration techniques.

Integrating over the parton distributions and final state momenta in general
requires performing a $(3N_{final}-4)+2$ dimensional integral
over a phase space which may include many cuts. Here $N_{final}$ is the 
number of particles in the final state. If the number of dimensions is
large, the integral is most easily carried out 
using Monte Carlo techniques.  Monte Carlo integration
approximates the integral by taking the average of a number of points, $N$,
selected at random, and multiplying by the volume, $V$, over which one is
integrating,
\begin{equation}
\int f(x) dx \simeq  {1 \over N} \sum_i f(x_i) \times V.
\end{equation}
Provided the function $f(x)$ which is to be integrated is sufficiently
flat, the number of
points for convergence is independent of the number of
dimensions. However, if $f(x)$ is sharply peaked convergence may
be exponentially slow.

In order to use Monte Carlo techniques for integrating a sharply
peaked function, it is necessary to remove the peaks. Peaks which are
analytically integrable, and for which the integral is invertible, can be
smoothed with the appropriate transformation of variables. A
Breit-Wigner resonance is an excellent example for such a case. The 
transformation
$y=\arctan(x)$ removes the peak and makes the integrand flat. Collinear
and soft poles often require more involved transformations which may
not have general analytical solutions.

Adaptive Monte Carlo programs such as VEGAS are able to flatten peaks
by using numeric approximations of the integrand. The result is not as
fast, or efficient as analytically removing the peaks, however it is
more convenient. 
For most applications this is very desirable. The major restriction is
that programs such as VEGAS can only remove peaks which are in the
plane of one of the integration variables. For example, these programs
will successfully flatten the peaks in the function 
\begin{equation}
f(x,y) = {1\over x}\, {1\over y}~, 
\end{equation}
but they will not be able to flatten those for 
\begin{equation}
f(x,y) = {1\over x+y}\, {1\over x-y}~,
\end{equation}
unless a change of variables is performed.
For processes with relatively few peaks, it is usually possible to map
each peak to one of the integration variables. For
complicated process such as $q\bar q\to\ell^+\ell^-\gamma\gamma$ and $q\bar
q'\to\ell\nu\gamma\gamma$, this is not the case.

In cases where it is not possible to simultaneously map every
peak to an integration variable, there are two classes of solutions
available. The first is to divide up phase space with cuts, such that
the peaks in each region can be mapped to the integration
variables. An adaptive Monte Carlo integration routine is used for 
each region separately. The results from each region are combined to obtain 
the total cross section. This method is effective for processes
with relatively simple peaking structure, however as the number of
peaks increases the technique quickly becomes cumbersome and prone to
error.

The second technique~\cite{excal,KP} is to choose points in phase space
not according to a single distribution, but according to the sum of
multiple distributions. Each distribution is responsible for a
specific set of peaks. 
The optimal number of points from each distribution is chosen using an
algorithm which minimizes the Monte Carlo integration error~\cite{KP}.
With each channel, a different set of poles is analytically
removed. The resulting code is very fast
and efficient, however it requires significant analytic work, which
must be repeated for each new process. 

In our approach, we have combined the power of
multichannel integration with the convenience offered by adaptive
Monte Carlo integration routines such as VEGAS. The result is a general
and flexible multiconfiguration Monte Carlo program called MCMC which can 
numerically integrate
sharply peaked functions in many dimensions with minimal input from
the user. 

Feynman diagrams offer a convenient mechanism for determining in which
dimensions peaks may appear. At tree level strong peaks in the cross
section are always associated with a propagator going on-shell. We
have implemented a general phase-space generator based on Feynman
diagrams.  Given a tree-level Feynman diagram, it maps a set of random
numbers to a point in phase space such that each propagator represents
one of the dimensions of integration.  The operation is invertible so
it can also return the set of random numbers associated with any point
in phase space. The user specifies the momentum flow of the contributing 
Feynman diagrams in a simple include file, together with the masses and
widths of the Breit-Wigner resonances which appear in each
diagram. All other aspects of the phase space integration are handled
automatically by the program. A more detailed description of the
approach will be given elsewhere~\cite{tim}. 

\begin{figure}
\phantom{x}
\vskip 14.cm
\includegraphics{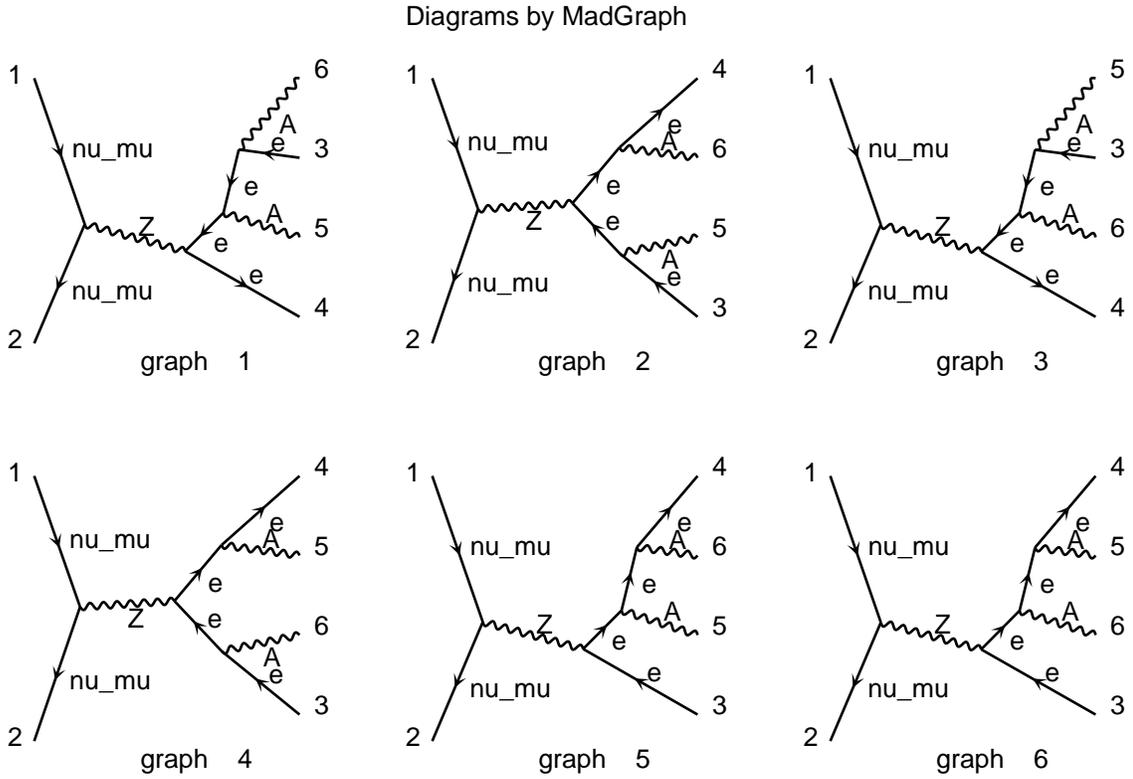}
\caption{The Feynman diagrams contributing to $\nu_\mu\bar\nu_\mu\to
e^+e^-\gamma\gamma$ at tree level, as generated by MadGraph. $A$ ($Z$)
represents a photon ($Z$ boson), $e$ an electron or positron, and
$nu\_ mu$ a muon neutrino, $\nu_\mu$.}
\label{fig:one}
\end{figure}

The convenience of MCMC is best illustrated in a simple
example. Consider the process $\nu_\mu \bar\nu_\mu \rightarrow e^+ e^-
\gamma\gamma$ for a center of mass energy of $\sqrt{s}=100$~GeV, where
each of the final state particles is required to have a
transverse momentum $p_T > 10$~GeV. The electron mass is assumed to be 
variable. The six Feynman diagrams associated
with this process can easily be generated with a program such as
MadGraph~\cite{madgraph}. The diagrams generated by MadGraph 
are shown in Fig.~\ref{fig:one}. Each diagram
represents a phase space configuration in the MCMC code with the 
appropriate poles mapped to the integration variables. 

These configurations are input to the integration package, which then
searches for peaks, and determines the optimal number of points to
choose from each configuration using an algorithm which minimizes the 
integration error. Table~\ref{tab:mass} compares the cross
sections obtained using the traditional single configuration approach 
with those from MCMC as the electron mass is varied from 0.01~GeV to
10~GeV. Notice that for large masses, the matrix element is relatively
flat and a single configuration accurately integrates the cross
section. However, as the electron mass decreases, the contribution from
the collinear regions becomes increasingly important, and the single 
configuration
package is unable to accurately integrate the function. Not only is
the error larger, but even with $5\times 10^6$ integration points, the single
configuration integration is giving the wrong result as it samples all of the
points from the peaks it has mapped to integration variables, 
completely neglecting the peaks which are only sampled by the other
configurations. Due to the mass singular terms associated with final
state radiation in the collinear limit, the cross section scales
approximately with $(\log(m_e^2/s))^2$ for small electron masses,
$m_e\leq 1$~GeV. This provides a simple check on the accuracy of the
MCMC result.

The primary difference between our approach and other
multichannel techniques is its generality. The mapping of
uniformly distributed random numbers to points in phase space can be
broken down into two steps.
First the uniformly distributed random numbers 
are deformed into non-uniform numbers, 
with a corresponding Jacobian. Next
these non-uniform numbers are mapped to four-momenta in phase space
with another Jacobian. For each step, one can choose to perform an
analytic transformation which will be very efficient for the
particular process being studied, or one can choose a general
transformation which is not optimized for the specific process, but
will work for any process. The approach of Ref.~\cite{KP} produces 
highly optimized
code for both transformations. In Ref.~\cite{Ohl}, a general 
procedure for the transformation from uniform space to a deformed
space is given, but optimized procedures for the
transformation to phase space are chosen. The MCMC program provides general
algorithms for both transformations, similar to the approach used by
CompHEP~\cite{comphep,ilyin} to perform the phase space integration. The 
resulting code is in general slightly slower than that resulting from 
the other two approaches, however we believe its user 
friendliness makes up for this short coming. The advantage of user
friendly programs at the expense of computer time has already been 
demonstrated by packages such as MadGraph, CompHEP~\cite{comphep,grace} and
GRACE~\cite{grace} which quickly produce
non-optimized tree-level matrix elements.  Indeed the synthesis of the
integration package outlined here with automatically generated matrix elements
will allow
the user to concentrate on the physics issues rather than numerical
integration techniques. While MCMC naturally interfaces with MadGraph
and HELAS~\cite{helas}, matrix elements resulting from any other
automated or non-automated calculation can be used. 

\section{$\ell\nu\gamma\gamma$ and $\ell^+\ell^-\gamma\gamma$ Production
at the Fermilab Tevatron}

We shall now discuss the calculation of $\ell\nu\gamma\gamma$ and
$\ell^+\ell^-\gamma\gamma$ production in hadronic collisions, together
with some phenomenological applications relevant for future $W$ mass
measurements at hadron colliders. 
To calculate the matrix elements for $q\bar
q\to\ell^+\ell^-\gamma\gamma$ and $q\bar q'\to\ell^\pm\nu\gamma\gamma$
we use MadGraph which automatically generates the SM matrix elements in
HELAS format. When photon exchange is taken into account, 40~Feynman
diagrams contribute to $\ell^+\ell^-\gamma\gamma$ production, while
there are 21~diagrams for $\ell^\pm\nu\gamma\gamma$ production. 
Taking into account symmetries in the phase space mapping,
20~(12) different configurations contribute to $q\bar 
q\to\ell^+\ell^-\gamma\gamma$ ($q\bar q'\to\ell\nu\gamma\gamma$).

In order to maintain electromagnetic gauge invariance for 
$q\bar q'\to\ell^\pm\nu\gamma\gamma$ in 
presence of finite $W$ width effects, the $W$ propagator and the
$WW\gamma$ and $WW\gamma\gamma$ vertex functions in the amplitudes
generated by MadGraph have to be modified~\cite{BHKSZ,BZ}. Finite
width effects are included by resumming the imaginary part of the $W$
vacuum polarization, $\Pi_W(q^2)$. The transverse part of $\Pi_W(q^2)$
receives an imaginary contribution 
\begin{equation}
{\rm Im}\,\Pi^T_W(q^2)=q^2{\Gamma_W\over M_W}
\end{equation}
while the imaginary part of the longitudinal piece vanishes. 
The $W$ propagator is thus given by
\begin{eqnarray}
\label{EQ:PROP}
D^{\mu\nu}_W(q)= \frac{-i}{q^2 - M^2_W + iq^2 \gamma_W}
\left[ g^{\mu\nu} - \frac{q^\mu q^\nu}{M^2_W}
( 1 + i \gamma_W ) \right],
\end{eqnarray}
with
\begin{equation}
\gamma_W={\Gamma_W\over M_W},
\end{equation}
where $\Gamma_W$ denotes the $W$ width.
A gauge invariant expression for the amplitude is then obtained by
attaching the final state photons to all charged particle propagators,
including those in the fermion loops which contribute to $\Pi_W(q^2)$.
As a result, the lowest order $WW\gamma$ and $WW\gamma\gamma$ vertex
functions, $\Gamma^{\alpha\beta\mu}_0$ and
$\Gamma^{\alpha\beta\mu\rho}_0$, are modified~\cite{BHKSZ,BZ} to
\begin{eqnarray}
\label{EQ:VERTEX}
\Gamma^{\alpha\beta\mu} &=& \Gamma^{\alpha\beta\mu}_0 ( 1 + i \gamma_W),
\\
\Gamma^{\alpha\beta\mu\rho} &=& \Gamma^{\alpha\beta\mu\rho}_0 ( 1 + 
i \gamma_W ).
\end{eqnarray}

The SM parameters used in our numerical calculations are $M_W=80.3$~GeV, 
$\Gamma_W=2.046$~GeV, $M_Z=91.19$~GeV, $\Gamma_Z=2.49$~GeV, and 
$\alpha(M_Z^2)=1/128$. These values are consistent
with recent measurements at LEP, LEP2, the SLC and the
Tevatron~\cite{lanc}. We use the parton distribution functions set~A of
Martin-Roberts-Stirling~\cite{mrsa} with the factorization scale set
equal to the parton center of mass energy $\sqrt{\hat s}$. All numerical
results are obtained for $p\bar p$ collisions with a center of mass
energy of $\sqrt{s}=1.8$~TeV. In Run~II, the Tevatron collider is foreseen to
operate at $\sqrt{s}=2$~TeV. For a center of mass energy of 2~TeV,
results qualitatively similar to those reported here are obtained. Cross
sections are about 5\% higher than those found for $\sqrt{s}=1.8$~TeV.
Since the total cross sections for $\ell^+\nu\gamma\gamma$ 
and $\ell^-\nu\gamma\gamma$ production are equal in $p\bar p$
collisions, we shall not consider the $\ell^-\nu\gamma\gamma$ channel
in the following.

To simulate the fiducial and kinematic acceptances of detectors, we
impose the following
transverse momentum ($p_T$) and pseudo-rapidity ($\eta$) cuts on
electrons and muons:
\begin{quasitable}
\begin{tabular}{cc}
electrons & muons\\
\tableline
$p_{T}^{}(e)         > 20$~GeV  & $p_{T}^{}(\mu)         > 25$~GeV\\
$|\eta(e)|           < 2.5$     & $|\eta(\mu)|           < 1.0$\\
$p\llap/_T           > 20$~GeV  & $p\llap/_T           > 25$~GeV\\
\end{tabular}
\end{quasitable}
Here, $p\llap/_T$ denotes the missing transverse momentum which we
identify with the transverse momentum of the neutrino in
$\ell\nu\gamma\gamma$ production. The $p\llap/_T$ cut is only applied in
$\ell\nu\gamma\gamma$ production. 
The cuts listed above approximately model the acceptance of the
CDF detector for electrons and muons in Run~I. Qualitatively similar
numerical results are obtained if cuts are used which approximate the
phase space region covered by the upgraded CDF detector for
Run~II~\cite{cdfrun2}, or if cuts are used which model the acceptance
of the D\O\ detector~\cite{D0upgr}.

In addition to the lepton cuts listed above, a pseudo-rapidity cut 
\begin{equation}
|\eta(\gamma)|<3.6,
\end{equation}
and a transverse momentum cut on the photons are imposed. 
In order to be able to accurately determine the shift in the $W$ and $Z$
boson masses induced by photon radiation correctly, it is necessary to
consider photon transverse momenta as low as the calorimeter threshold of the
detector, which is about 100~MeV. Subsequently, we therefore require 
\begin{equation}
p_T(\gamma)>0.1~{\rm GeV}
\end{equation}
in all our calculations unless stated otherwise explicitly. The photon 
transverse momentum and pseudo-rapidity cuts are necessary to avoid 
soft singularities and collinear divergences associated with initial 
state radiation. 

Since we are mostly interested in photon radiation in $W$ and $Z$
decays, we impose additional cuts on the di-lepton invariant mass,
\begin{equation}
75~{\rm GeV}<m(\ell\ell)<105~{\rm GeV,}
\label{eq:ll}
\end{equation}
and the transverse mass of the $\ell\nu$ system,
\begin{equation}
65~{\rm GeV}<m_T(\ell p\llap/_T)<100~{\rm GeV}.
\label{eq:ln}
\end{equation}
CDF and D\O\ utilize similar cuts in their $W$ mass 
analyses~\cite{cdfwmass,D0Wmass}.
Events satisfying Eqs.~(\ref{eq:ll}) and~(\ref{eq:ln}) are called
$Z\to\ell^+\ell^-$ and $W\to\ell\nu$ events, respectively, in the
following. 

\begin{figure}
\phantom{x}
\vskip 15cm
\includegraphics{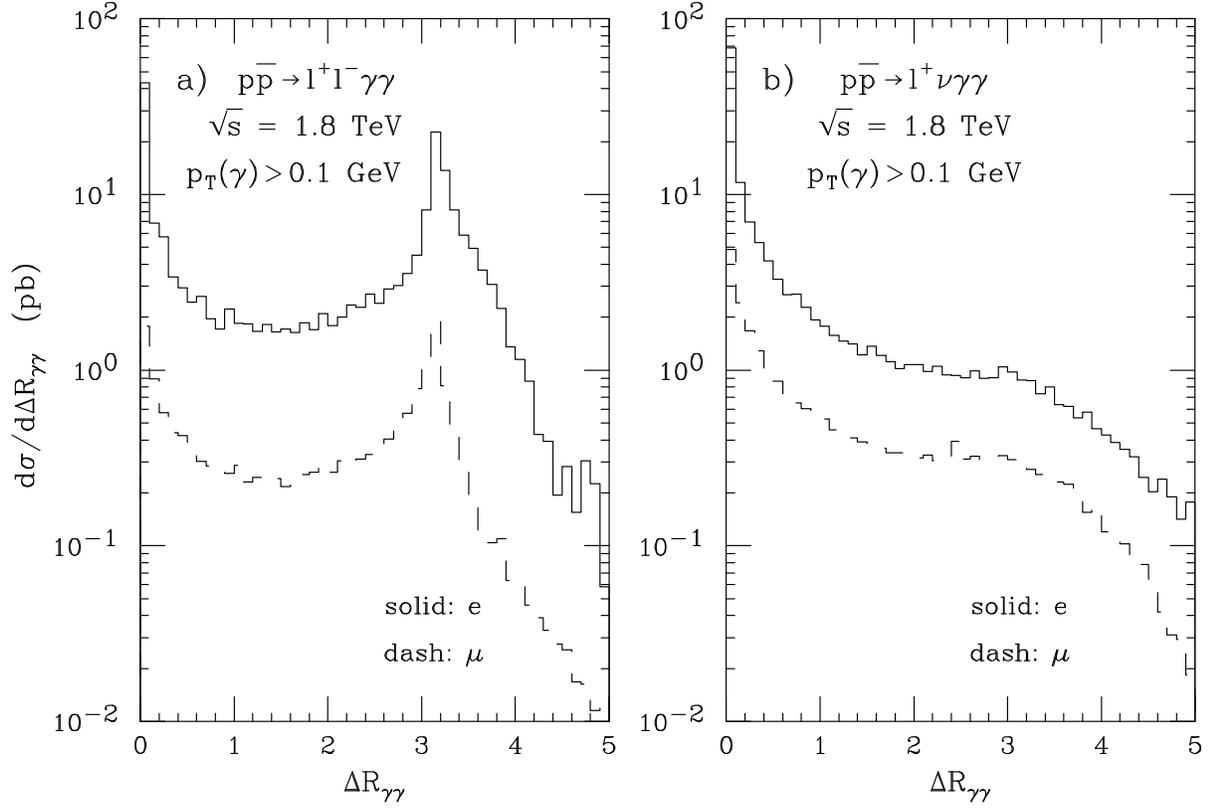}
\caption{The distribution of the separation between the two photons, $\Delta
R_{\gamma\gamma}$, for a) $p\bar p\to\ell^+\ell^-\gamma\gamma$ and b)
$p\bar p\to\ell^+\nu\gamma\gamma$ at $\sqrt{s}=1.8$~TeV. The solid
and dashed histograms show the differential cross sections for electrons
and muons, respectively. The cuts imposed are described in the text.}
\label{fig:two}
\end{figure}

To demonstrate that the multiconfiguration Monte Carlo approach we use yields
accurate results both in the collinear region as well as for photons
emitted at large angles, we show in Fig.~\ref{fig:two} the differential 
cross section versus the separation between the two photons in the azimuthal 
angle-pseudorapidity plane,
\begin{equation}
\Delta R_{\gamma\gamma}=\sqrt{\Delta\phi_{\gamma\gamma}^2 +
\Delta\eta_{\gamma\gamma}^2}~.
\end{equation}
The strong
peak for small $\Delta R_{\gamma\gamma}$ arises when both photons are
emitted by the same charged lepton, and the photons are collinear with
the lepton. The peak at $\Delta R_{\gamma\gamma}\approx 3$ in
$\ell^+\ell^-\gamma\gamma$ production originates from Feynman diagrams
where the photons are radiated off different leptons. Since photons do
not couple to neutrinos, this peak is absent in $\ell\nu\gamma\gamma$
production. For electrons, the collinear peaks are significantly more
pronounced than for muons. The difference in the differential cross
sections for electrons and muons away from the collinear regions is
entirely due to the different $p_T$ and rapidity cuts imposed on these
particles. The statistical fluctuations are quite uniform over the
full range of $\Delta R_{\gamma\gamma}$ values considered, indicating
that the MCMC program distributes the generated events and their weights
appropriately. Away from the collinear peaks, our 
calculation\footnote{Parton level FORTRAN programs for
$p\,p\hskip-7pt\hbox{$^{^{(\!-\!)}}$} \to\ell^\pm\nu\gamma\gamma$ and 
$p\,p\hskip-7pt\hbox{$^{^{(\!-\!)}}$} \to\ell^+\ell^-\gamma\gamma$
which include the MCMC source code are available upon request from the
authors.} agrees
with that of Ref.~\cite{BHKSZ} to better than 1\%. In this region,
conventional adaptive Monte Carlo routines such as VEGAS are sufficient
in order to obtain a numerically stable result. The distributions of
the separation between the photons and the charged lepton (leptons) are
qualitatively very similar to the $\Delta R_{\gamma\gamma}$ spectrum.

In Tables~\ref{tab:two} and~\ref{tab:three}, we list the fraction of
$W\to\ell\nu$ and $Z\to\ell^+\ell^-$ events at the Tevatron which 
contain two photons as a function of the minimum photon transverse
momentum. For comparison, we also list the event fractions containing
one photon. Fractions are obtained by normalization with respect to the
lowest order cross section within cuts. The results for $\ell\nu\gamma$ 
and $\ell^+\ell^-\gamma$ production are obtained using the
calculation of Ref.~\cite{BZ}. No lepton-photon or photon-photon
separation cuts are imposed.

Approximately 3\% (1\%) of all $W\to e\nu$ ($W\to\mu\nu$) events, and
14\% (5\%) of all $Z\to e^+e^-$ ($Z\to\mu^+\mu^-$) events, contain
two photons with a minimum transverse momentum of $p_T^{min}(\gamma)=0.
1$~GeV. Because of the mass singular logarithms associated with 
final state photon bremsstrahlung in the collinear limit, the fraction
of $W\to e\nu$ and $Z\to e^+e^-$ 
events with two photons is more than a factor~3 larger than the
corresponding fraction of $W\to\mu\nu$ and $Z\to\mu^+\mu^-$ events.
In contrast to $W$ events, both leptons can radiate photons in $Z$
decays. As a result, the probability of $Z\to\ell^+\ell^-$ events to
radiate two photons is more than four times that of
$W\to\ell\nu$ events. For increasing $p_T^{min}(\gamma)$, the fraction of
$W$ and $Z$ events containing photons drops quickly. 

For small photon transverse momenta, the cross section is completely
dominated by final state radiation. In this region, the fraction of
events containing two photons, $P_2$, can be estimated using the simple 
formula~\cite{multi}
\begin{equation}
P_2={P_1^2\over 2}~,
\label{eq:naive}
\end{equation}
where $P_1$ is the fraction of events containing one photon. The results
obtained using Eq.~(\ref{eq:naive}) are also listed in 
Tables~\ref{tab:two} and~\ref{tab:three}. For large values of
$p_T(\gamma)$, the available phase space for final state radiation is
strongly reduced by the transverse momentum cuts imposed on the leptons,
and initial state radiation plays an increasingly important role. For
$p_T^{min}(\gamma)\geq 3$~GeV, Eq.~(\ref{eq:naive}) therefore becomes
more and more inaccurate. 

\begin{figure}
\phantom{x}
\vskip 15cm
\includegraphics{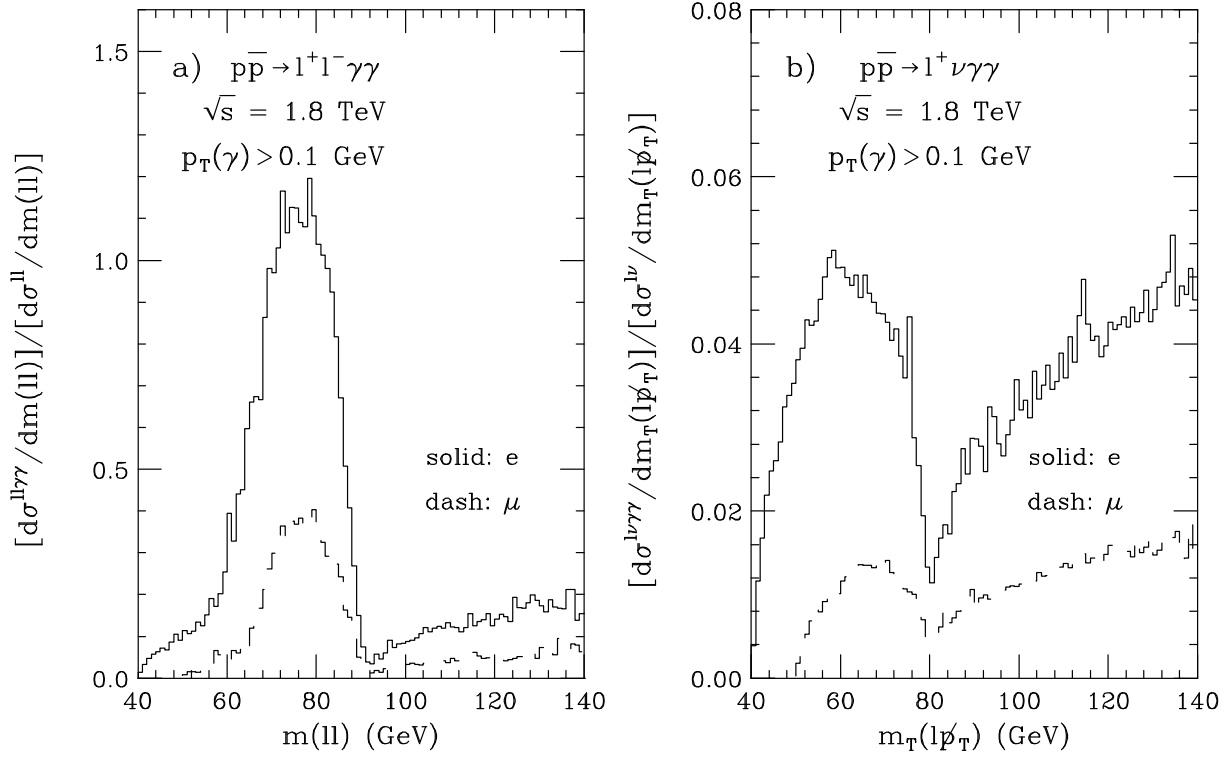}
\caption{Ratio of a) the $p\bar p\to\ell^+\ell^-\gamma\gamma$ and the
lowest order $p\bar p\to\ell^+\ell^-$ cross sections as a function of 
the $\ell^+\ell^-$ invariant mass, and b) the $p\bar 
p\to\ell^+\nu\gamma\gamma$ and the lowest order $p\bar p\to\ell^+\nu$
cross section versus $m_T(\ell p\llap/_T)$ at $\sqrt{s}=1.8$~TeV. The solid
and dashed histograms show the cross section ratios for electrons
and muons, respectively. The cuts imposed are described in the text.}
\label{fig:three}
\end{figure}

The large mass singular terms associated with final state bremsstrahlung
result in a significant change in the shape of the $m_T(\ell
p\llap/_T)$ and the di-lepton invariant mass distributions. This is 
demonstrated in Fig.~\ref{fig:three}. Here we do not impose the 
di-lepton invariant mass cut and the $\ell\nu$ transverse mass cut of 
Eqs.~(\ref{eq:ll}) and~(\ref{eq:ln}). In Fig.~\ref{fig:three}a we show 
the ratio of the
$\ell^+\ell^-\gamma\gamma$ and the lowest order $\ell^+\ell^-$ cross
section as a function of $m(\ell\ell)$. The cross section ratio is 
seen to vary rapidly. The dip at $m(\ell\ell)=M_Z$ is a direct 
consequence of the Breit-Wigner resonance of the $Z$ boson. Below the
$Z$ peak, the cross section
ratio rises very sharply and in the region $70~{\rm GeV}<m(ee)<80~{\rm
GeV}$, the cross section ratio is of order one in the electron case.
The dip located at $m(\ell\ell)=M_Z$ and the substantially enhanced 
rate of events with two photons below the resonance peaks are caused by 
final state bremsstrahlung in events where the $\ell^+\ell^-\gamma\gamma$
invariant mass is close to $M_Z$. 

Fig.~\ref{fig:three}b displays
the ratio of the $\ell^+\nu\gamma\gamma$ and the $\ell^+\nu$ cross
section as a function of the $\ell\nu$ transverse mass. Here the dip at
$m_T(\ell p\llap/_T)=M_W$ is due to the Jacobian peak in the $\ell\nu$
transverse mass distribution. Because of the long tail of the lowest
order $m_T(\ell p\llap/_T)$ distribution below $M_W$ and the fact that 
photons are not radiated by neutrinos, the enhancement in the
$\ell\nu\gamma\gamma$ to $\ell\nu$ cross section ratio is less
pronounced than that encountered in the $\ell^+\ell^-$ case. In the
region of large transverse masses, $m_T(\ell p\llap/_T)>100$~GeV, the
shape of the transverse mass distribution is sensitive to the $W$
width. Fig.~\ref{fig:three}b shows that two photon radiation
significantly modifies the shape of the $m_T(\ell p\llap/_T)$
distribution in this region. This will directly influence the $W$ width
extracted by experiment.

The shape changes in the $\ell\nu$ transverse mass and the di-lepton
invariant mass distributions suggest that two photon radiation may have 
a non-negligible effect on the measured $W$ and $Z$ masses, and also on
the $W$ width extracted from the high transverse mass region. Since the 
shape change caused by two photon radiation in the
distribution used to extract the mass is more pronounced in the $Z$ case,
the shift in the $Z$ boson mass is expected to be
considerably larger than the shift in $M_W$. For a 
realistic calculation of how ${\cal O}(\alpha^2)$ corrections affect the 
$W$ and $Z$ resonance parameters, soft and virtual corrections and 
detector resolution effects need to be included. 

We have not taken into account detector resolution effects or realistic 
lepton and photon
identification requirements in the calculations presented in this
Section. In particular, we have
assumed that photons and leptons with arbitrary small opening angles can
be discriminated. In practice, the finite resolution of the
electromagnetic calorimeter makes 
it difficult to separate electrons and photons for small opening angles
between their momentum vectors. Electron and photon four-momentum
vectors are therefore recombined if their separation in the azimuthal
angle-pseudorapidity plane is smaller than a critical 
value~\cite{cdfwmass,D0Wmass}. This
eliminates the mass singular terms associated with final state photon
radiation and thus may reduce the fraction of $W$ and $Z$ events with
two photons significantly. Since muons are identified by hits in the
muon chambers, the four momentum vectors of muons and photons are not
combined for small opening angles. Instead, one frequently requires the
photon energy to be below a threshold $E_c$ in a cone around the
muon. The mass singular logarithms thus survive in the muon case.
The precise lepton identification requirements and their effects on the
size of the EW corrections $W$ and $Z$ boson production are detector 
dependent. 

\section{Summary and Conclusions}

The mass of the $W$ boson is one of the fundamental parameters of the SM
and a precise measurement of $M_W$ is an important objective for current
experiments at LEP2 and future experiments at the Tevatron. A precise
measurement of $M_W$ helps to constrain the Higgs boson mass from
radiative corrections. It will also provide restrictions on the
parameters of the MSSM. In order to perform such a measurement at a
hadron collider, it is crucial to fully control higher order QCD and
EW corrections to $W$ production. In a precision measurement of
$M_W$ in hadronic collisions, a simultaneous determination of the mass
of the $Z$ boson is required for calibration purposes. A detailed
understanding of the QCD and electroweak corrections to $Z$ boson
production is therefore also necessary.

Recent calculations~\cite{BKW,BKS} have shown that the ${\cal
O}(\alpha)$ electroweak 
corrections to $W$ and $Z$ production have a significant impact on the
weak boson masses extracted from experiment. The dominant contribution
originates from final state photon radiation. The magnitude of the shift
in $M_W$ and $M_Z$ induced by the ${\cal O}(\alpha)$ corrections
suggests that ${\cal O}(\alpha^2)$ corrections may have an effect which
cannot be ignored in future $W$ mass measurements at the Tevatron. In
this paper we have presented a calculation of the real ${\cal
O}(\alpha^2)$ photonic corrections to $W$ and $Z$ boson production in
hadronic collisions. Our calculation is based on the full set of Feynman
diagrams contributing to $\ell^+\ell^-\gamma\gamma$ and
$\ell\nu\gamma\gamma$ production and includes finite lepton mass
effects. In order to maintain gauge invariance in $\ell\nu\gamma\gamma$
production, the $W$ propagator and the $WW\gamma$ and $WW\gamma\gamma$
vertex functions are modified using the prescription given in
Refs.~\cite{BHKSZ} and~\cite{BZ}. 

In order to accurately determine the shift in the $W$ and $Z$
masses caused by photon radiation, the numerical calculation should be
stable for arbitrarily small or large lepton - photon opening angles as 
well as for photon energies as small as the tower threshold of the
electromagnetic calorimeter of the Tevatron experiments, which is of
${\cal O}(100~{\rm MeV})$. Due to the collinear and soft singularities
present, this poses a challenge. Standard adaptive Monte Carlo integration
routines such as VEGAS do not yield a stable result for processes with a
complicated peaking structure in the matrix elements, such as
$q\bar q\to\ell^+\ell^-\gamma\gamma$ and $q\bar
q'\to\ell\nu\gamma\gamma$. To obtain numerically stable and accurate 
results in these cases, multi-channel Monte Carlo
integration techniques are frequently used. This approach requires that
the peaks in the matrix elements are analytically mapped. To calculate
the cross sections for 
$\ell^+\ell^-\gamma\gamma$ and $\ell\nu\gamma\gamma$ production at
hadron colliders, we developed a multiconfiguration Monte Carlo 
integration routine called MCMC which is based on a similar approach,
adding the benefit of largely automizing the mapping of the peaks. MCMC
thus can be used to calculate other processes with matrix elements
exhibiting a complex set of peaks with almost no additional effort. The
algorithm which is used in MCMC to map out the peaks is based on the
Feynman diagrams which contribute to the process considered. 

Imposing $W$ and $Z$ boson selection cuts on the final state leptons, we 
found that a significant fraction of weak boson events contains two
photons. The probability for $Z$ events to radiate two photons is almost
a factor five larger than that for $W$ events. For $W\to\mu\nu$ and
$Z\to\mu^+\mu^-$ decays, the rate for two 
photon radiation is about a factor~3 smaller than the corresponding rate 
for decays with electrons in the final state.
If the photon $p_T$ is less than about 3~GeV, the fraction of $W$
and $Z$ events containing two photons can be estimated with an accuracy
of 20\% or better using a simple equation (see Eq.~(\ref{eq:naive})). 

Two photon radiation was also found to significantly alter the shapes
of the $Z$ boson resonance curve and the $\ell\nu$ transverse mass
distribution. The shift in the $W$ and $Z$ masses, and in the $W$ width
measured from the tail of the transverse mass distribution, 
caused by the ${\cal
O}(\alpha^2)$ real photon corrections may thus be non-negligible for
future hadron collider experiments. For a realistic estimate of how
strongly the ${\cal O}(\alpha^2)$ corrections affect the $W$ boson
parameters extracted from experiment it is necessary to include the
effects of soft and virtual corrections, as well as detector resolution
effects. The calculation of
$\ell^+\ell^-\gamma\gamma$ and $\ell\nu\gamma\gamma$ production
presented in this paper thus only is the first step towards a more
complete understanding of the ${\cal O}(\alpha^2)$ electroweak
corrections to $W$ and $Z$ production in hadronic collisions.

\acknowledgements
We would like to thank R.~Brock, Y-K.~Kim, M.~Lancaster, D.~Waters and
D.~Wood for stimulating discussions.
One of us (U.B.) is grateful to the Fermilab Theory Group,
where part of this work was carried out, for its generous hospitality.
This work has been supported in part by DOE
contract No.~DE-FG02-91ER40677 and NSF grant~PHY-9600770. 

%
%

%
\newpage
%
\widetext
\begin{table}
\caption{Integrated cross section for the process $\nu_\mu \bar\nu_\mu
\rightarrow e^+ e^- \gamma \gamma$ at $\sqrt{s}=100$~GeV as a function
of the electron mass
for single and multi configuration adaptive Monte Carlo integration. A
$p_T>10$~GeV cut is imposed on all final state particles. In
all cases $8\times 100,000$ events are generated to set the grid, and 
$5\times 1$~million events for evaluating the integral.} 
\label{tab:mass}
\vskip 5.mm
\begin{tabular}{ccccc}
 electron mass & \multicolumn{2}{c}{Multiconfiguration} & 
\multicolumn{2}{c}{Single configuration} \\
(GeV) & $\sigma$ (fb) & $\chi^2$ & $\sigma$ (fb) & $\chi^2$ \\
\tableline
 0.01&  88.4  $\pm$ 0.8  & 0.2   & 48.9  $\pm$ 3     & 0.3\\
 0.1&   46.34 $\pm$ 0.08 & 0.7   & 36.5  $\pm$ 2     & 1.1\\
 1.0&   17.14 $\pm$ 0.03 & 1.    & 16.5  $\pm$ 0.2   & 0.7\\
10.0&   1.999 $\pm$ 0.003& 0.3   & 1.992 $\pm$ 0.003 & 0.2  \\
\end{tabular}
\end{table}

%
\newpage
\begin{table}
\caption{Fraction of $W\to e\nu$ and $W\to\mu\nu$ events (in percent) 
containing one
or two photons with a transverse momentum $p_T(\gamma)>p_T^{min}(\gamma)$
at the Tevatron ($p\bar p$ collisions at $\sqrt{s}=1.8$~TeV). Fractions 
are obtained with respect
to the lowest order cross section. The cuts imposed are specified in the
text. The relative statistical error on the event fractions from the 
Monte Carlo integration is approximately 1\%.} 
\label{tab:two}
\vskip 5.mm
\begin{tabular}{cccc}
$p_T^{min}(\gamma)$ (GeV) & $W\to e\nu\gamma$ & $W\to e\nu\gamma\gamma$
& $W\to e\nu\gamma\gamma$ [Eq.~(\ref{eq:naive})] \\
\tableline
0.1 & 23.9 & 3.05 & 2.86 \\
0.3 & 17.3 & 1.56 & 1.50 \\
1 & 10.4 & 0.53 & 0.54 \\
3 & 4.82 & 0.09 & 0.12 \\
10 & 0.56 & $1.3\times 10^{-3}$ & $1.6\times 10^{-3}$ \\
\tableline
\tableline
$p_T^{min}(\gamma)$ (GeV) & $W\to\mu\nu\gamma$ & $W\to\mu\nu\gamma\gamma$
& $W\to\mu\nu\gamma\gamma$ [Eq.~(\ref{eq:naive})] \\
\tableline
0.1 & 13.5 & 0.99 & 0.91 \\
0.3 & 9.65 & 0.48 & 0.47 \\
1 & 5.74 & 0.17 & 0.16 \\
3 & 2.65 & $2.7\times 10^{-2}$ & $3.5\times 10^{-2}$ \\
10 & 0.33 & $6.3\times 10^{-4}$ & $5.4\times 10^{-4}$ \\
\end{tabular}
\end{table}

\newpage
\begin{table}
\caption{Fraction of $Z\to e^+e^-$ and $Z\to\mu^+\mu^-$ events (in percent) 
containing one
or two photons with a transverse momentum $p_T(\gamma)>p_T^{min}(\gamma)$
at the Tevatron ($p\bar p$ collisions at $\sqrt{s}=1.8$~TeV). Fractions 
are obtained with respect
to the lowest order cross section. The cuts imposed are specified in the
text. The relative statistical error on the event fractions from the 
Monte Carlo integration is approximately 1\%.} 
\label{tab:three}
\vskip 5.mm
\begin{tabular}{cccc}
$p_T^{min}(\gamma)$ (GeV) & $Z\to e^+e^-\gamma$ & $Z\to e^+e^-\gamma\gamma$
& $Z\to e^+e^-\gamma\gamma$ [Eq.~(\ref{eq:naive})] \\
\tableline
0.1 & 52.3 & 14.6 & 13.7 \\
0.3 & 39.1 & 7.80 & 7.64 \\
1 & 25.0 & 3.05 & 3.13 \\
3 & 12.9 & 0.61 & 0.83 \\
10 & 2.17 & $1.3\times 10^{-2}$ & $2.4\times 10^{-2}$ \\
\tableline
\tableline
$p_T^{min}(\gamma)$ (GeV) & $Z\to\mu^+\mu^-\gamma$ & 
$Z\to\mu^+\mu^-\gamma\gamma$ & $Z\to\mu^+\mu^-\gamma\gamma$ 
[Eq.~(\ref{eq:naive})] \\
\tableline
0.1 & 31.2 & 4.76 & 4.87 \\
0.3 & 23.8 & 2.67 & 2.83 \\
1 & 15.4 & 1.07 & 1.19 \\
3 & 8.19 & 0.25 & 0.34 \\
10 & 1.62 & $8.4\times 10^{-3}$ & $1.3\times 10^{-2}$ \\
\end{tabular}
\end{table}

\newpage
%
%
%
\newpage
%
%
%
\newpage
%
%
%
\end{document}